\documentclass[12pt,preprint]{aastex}

\begin{document}

\title{Formation of Giant Planets by Disk Instability on Wide Orbits 
Around Protostars with Varied Masses}

\author{Alan P.~Boss}
\affil{Department of Terrestrial Magnetism, Carnegie Institution of
Washington, 5241 Broad Branch Road, NW, Washington, DC 20015-1305}
\authoremail{boss@dtm.ciw.edu}

\begin{abstract}

Doppler surveys have shown that more massive stars have significantly 
higher frequencies of giant planets inside $\sim$ 3 AU than lower 
mass stars, consistent with giant planet formation by core accretion.
Direct imaging searches have begun to discover significant numbers
of giant planet candidates around stars with masses of $\sim$ 1 $M_\odot$ 
to $\sim$ 2 $M_\odot$ at orbital distances of $\sim$ 20 AU to $\sim$ 120 AU.
Given the inability of core accretion to form giant planets at such large 
distances, gravitational instabilities of the gas disk leading to clump
formation have been suggested as the more likely formation mechanism.
Here we present five new models of the evolution of disks with inner radii 
of 20 AU and outer radii of 60 AU, for central protostars with masses of 
0.1, 0.5, 1.0, 1.5, and 2.0 $M_\odot$, in order to assess the likelihood 
of planet formation on wide orbits around stars with varied masses. The
disk masses range from 0.028 $M_\odot$ to 0.21 $M_\odot$, with initial
Toomre $Q$ stability values ranging from 1.1 in the inner disks to
$\sim 1.6$ in the outer disks. These five models show that disk instability 
is capable of forming clumps on time scales of $\sim 10^3$ yr that, if they 
survive for longer times, could form giant planets initially on orbits with 
semimajor axes of $\sim$ 30 AU to $\sim$ 70 AU and eccenticities of $\sim$ 0 to $\sim$ 0.35, with initial masses of $\sim 1 M_{Jup}$ to $\sim 5 M_{Jup}$, around solar-type stars, with more protoplanets forming as the mass of the protostar (and protoplanetary disk) are increased. In particular,
disk instability appears to be a likely formation mechanism for the HR 8799
gas giant planetary system. 

\end{abstract}


\section{Introduction}

 Direct imaging searches for extrasolar planets have typically placed
only upper limits on the frequency of giant planets on orbits between 
$\sim 20$ AU and $\sim 100$ AU (Nielson et al. 2008; Nielsen
\& Close 2010) or $\sim 40$ AU and  
$\sim 200$ AU (Lafreni\'ere et al. 2007), for planets with masses above
4 $M_{Jup}$ or 2 $M_{Jup}$, respectively. These surveys led to upper
limits on the frequency of giant planet companions on such orbits
of $\sim$ 10\% to $\sim$ 20\%. These upper limits are, however, comparable
to estimates of the frequency of detected giant planets on orbits inside 
$\sim$ 3 AU of FGKM dwarfs by Doppler spectroscopy (Cumming et al. 2008).
Gravitational microlensing detections of ice and gas giant planets
orbiting beyond 3 AU imply an even higher frequency of planets,
about 35\% (Gould et al. 2010). Hence, significant numbers of 
giant planets on wide orbits might very well exist.

 Recently, persuasive evidence has begun to appear that wide giant planets 
do indeed exist in significant numbers. The A3V star (2.06 $M_\odot$)
Fomelhaut appears to have a planetary companion 119 AU away with a
mass less than 3 $M_{Jup}$, based on the planet's failure to disrupt 
the cold dust belt in which it is embedded (Kalas et al. 2008).
The A5V star (1.5 $M_\odot$) HR 8799 appears to have a system of
at least four gas giant planets, orbiting at projected distances 
of 14, 24, 38, and 68 AU, with minimum masses of 7, 7, 7, and 5 $M_{Jup}$, 
respectively, based on their luminosities and an estimated age of 
the system of 30 Myr (Marois et al. 2008, 2010).
The four HR 8799 exoplanets are also embedded in a dust debris disk 
(Su et al. 2009). A companion with a mass in the range of 10 to 40
$M_{Jup}$ has been detected at a projected separation of 29 AU
from the G9 star (0.97 $M_\odot$) GJ 758 (Thalmann et al. 2009), and
other good candidates for wide planetary companions have been proposed
as well (e.g., Oppenheimer et al. 2008; Lafreni\'ere, Jayawardhana, 
\& van Kerkwijk 2008). Heinze et al. (2010a,b), however, estimate
that no more than 8.1\% of the 54 sun-like stars studied in their
planet imaging survey could have planets similar to those of HR 8799.

 Doppler surveys have been extended to a range of stellar masses, providing 
the first estimates of how the planetary census depends on stellar type. 
A-type stars appear to have a significantly higher frequency of giant planets 
with orbits inside 3 AU compared to solar-type stars (Bowler et al. 2010).
M dwarfs, on the other hand, appear to have a significantly lower frequency 
of giant planets inside 2.5 AU than FGK dwarfs (Johnson et al. 2010).
Thus there is a clear indication that the frequency of giant planets
increases with stellar mass, at least for relatively short period orbits.
Assuming that protoplanetary disk masses increase with increasing
stellar mass, such a correlation is consistent with the core
accretion mechanism for giant planet formation, as a higher surface
density of solids leads to proportionately larger mass cores that
could become gas giant planets (e.g., Wetherill 1996; Ida \& Lin 2005).
However, microlensing detections (Gould et al. 2010) imply a considerably 
higher frequency ($\sim$ 35\%) of giant planets around early M dwarf stars 
($\sim 0.5 M_\odot$) than that found by the Doppler surveys, again orbiting at
larger distances than those probed by the Doppler surveys. Evidently
even early M dwarfs might also have a significant population of relatively
wide gas giant planets.

 Core accretion is unable, however, to form massive planets beyond $\sim$ 35 
AU, even in the most favorable circumstances (e.g., Levison \& Stewart 2001; 
Thommes, Duncan, \& Levison 2002; Chambers 2006), and gravitational
scattering outward appears to be unable to lead to stable wide orbits 
(Dodson-Robinson et al. 2009; Raymond, Armitage, \& Gorelick 2010). 
Disk instability (Boss 1997) is then 
the remaining candidate mechanism for forming wide gas giant planets 
(Dodson-Robinson et al. 2009; Boley 2009). Previous models found that disk 
instability could readily produce giant planets at distances of 20 AU to 30 
AU (Boss 2003), but not at distances of 100 AU to 200 AU (Boss 2006a). Here 
we present results for intermediate-size disks (20 AU to 60 AU) for
a range of central protostar masses (0.1 to 2.0 $M_\odot$), to learn
if the disk instability mechanism for giant planet formation is consistent 
with the results of the Doppler and direct imaging surveys to date.

\section{Numerical Methods}

 The calculations were performed with a numerical code that solves the three 
dimensional equations of hydrodynamics and radiative transfer in the 
diffusion approximation, as well as the Poisson equation for the 
gravitational potential. This same basic code has been 
used in all of the author's previous studies of disk instability. The code 
is second-order-accurate in both space and time. A complete description
of the entire code, including hydrodynamics and radiative transfer, may be 
found in Boss \& Myhill (1992), with the following exceptions: The 
central protostar is assumed to move in such a way as to preserve the location 
of the center of mass of the entire system (Boss 1998), which is 
accomplished by altering the location of the point mass source of the 
star's gravitational potential to balance the center of mass of the disk.
The Pollack et al. (1994) Rosseland mean opacities are used for the dust 
grains that dominate the opacities in these models. The energy
equation of state in use since 1989 is described by Boss (2007).
A flux-limiter for the diffusion approximation radiative transfer
was not employed, as it appears to have only a modest effect on
midplane temperatures (Boss 2008). Recent tests of the radiative 
transfer scheme are described in Boss (2009).

 The equations are solved on a spherical coordinate grid with $N_r = 101$
(including the central grid cell, which contains the central protostar),
$N_\theta = 23$ in $\pi/2 \ge \theta \ge 0$, and $N_\phi = 256$, with
$N_\phi$ being increased to 512 once fragments begin forming.
The radial grid is uniformly spaced with $\Delta r = 0.4$ AU between 20 
and 60 AU. The $\theta$ grid is compressed into the midplane to ensure 
adequate vertical resolution ($\Delta \theta = 0.3^o$ at the midplane). The 
$\phi$ grid is uniformly spaced. The number of terms in the spherical harmonic 
expansion for the gravitational potential of the disk is $N_{Ylm} = 32$
when $N_\phi = 256$, while $N_{Ylm} = 48$ when $N_\phi = 512$.

 The Jeans length criterion (e.g., Boss et al. 2000) and the Toomre
length criterion (Nelson 2006) are both monitored throughout
the evolutions to ensure that any clumps that might form are not 
numerical artifacts. The Jeans length criterion consists of
requiring that all of the grid spacings in the spherical coordinate
grid remain smaller than 1/4 of the Jeans length $\lambda_J =
\sqrt{ \pi c_s^2 \over G \rho}$, where $c_s$ is the local sound
speed, $G$ the gravitational constant, and $\rho$ the density.
Similarly, the Toomre length criterion consists of requiring that 
all of the grid spacings remain smaller than 1/4 of the Toomre 
length $\lambda_T = (2 c_s^2 / G \Sigma)$, where $\Sigma$ is the 
mass surface density. Once well-defined fragments form, these
criteria may be violated at the maximum densities of the clumps,
due to the non-adaptive nature of the spherical coordinate grid, 
as is expected to be the case for self-gravitating clumps that
are trying to contract to higher densities on a fixed grid.
However, provided that the Jeans and Toomre constraints are
satisfied at the time that well-defined clumps appear, these
clumps are expected to be genuine and not spurious artifacts.

 The boundary conditions are chosen at both 20 and 60 AU to absorb radial
velocity perturbations, to simulate the continued existence of the 
disk inside and outside the active numerical grid. As discussed
in detail by Boss (1998), the use of such non-reflective boundary
conditions should err on the side of caution regarding the growth
of perturbations, as found by Adams, Ruden, \& Shu (1989).
Mass and momentum that enters the innermost 
shell of cells at 20 AU are added to the central protostar, whereas 
mass or momentum that reaches the outermost shell of cells at 60 AU 
remains on the active hydrodynamical grid.

 The controversy over whether or not disk instability can lead to
protoplanet formation inside about 20 AU continues unabated (see
the recent reviews by Durisen et al. 2007 and Mayer, Boss, \& Nelson
2010). Attempts to find a single reason for different numerical outcomes
for disk instability models have been largely unsuccessful to date
(e.g., Boss 2007, 2008), implying that the reason cannot be traced to a 
single code difference, but rather to the totality of differences, such 
as spatial resolution, gravitational potential accuracy, artificial
viscosity, stellar irradiation effects, radiative transfer, numerical
heating, equations of state, initial density and temperature profiles,
disk surface boundary conditions, and time step size, to name a few.

 Comparison calculations on nearly identical disk models have led
Boss (2007) and Cai et al. (2010) to reach different conclusions.
While Boss (2007) concluded that fragmentation was possible inside 20 AU, 
Cai et al. (2010) found no evidence for fragmentation in their models, 
which included numerous improvements over their previous work, such as
a better treatment of radiative transfer in optically thin regions
of the disk and elimination of the spurious numerical heating in
the inner disk regions where Boss (2007) found fragments to form.
Cai et al. (2010) suggested that the main difference might be 
artificially fast cooling in the Boss models as a result of the 
thermal bath boundary conditions used in Boss models, which could
not be duplicated with the Cai et al. (2010) code because of numerical
stability problems. Analytical test cases have been advanced as
one means for testing radiative transfer in the numerical codes 
(e.g., Boley et al. 2006). Boss (2009) derived two new analytical
radiative transfer solutions and showed that Boss code does an 
excellent job of handling the radiative boundary conditions of a disk 
immersed in a thermal bath; the Boss code relaxes to the analytical 
solutions for both a spherically symmetric cloud and an axisymmetric disk.

 Recently, Boss (2010) published models showing that disk instability is considerably less robust inside 20 AU in disks with half the mass of 
previous models (e.g., Boss 2007), but still possible. Inutsuka, 
Machida, \& Matsumoto (2010) found in their magnetohydrodynamic
collapse calculations that the massive disks that formed were subject
to gravitational instability and fragment formation, even inside 
20 AU. Arguments against inner disk fragmentation are often based on 
simple cooling time estimates (e.g., Cai et al. 2010). However,  
Meru \& Bate (2010, 2011) have emphasized that many previous numerical
calculations with fixed cooling times are likely to have reached
incorrect results, in part as a result of insufficient spatial resolution.
Meru \& Bate (2010, 2011) presented numerous disk instability models 
that underwent fragmentation inside 20 AU for a variety of initial conditions.
While the debate over inner disk fragmentation is likely to continue,
the present models should be considerably less controversial, given
their restriction to fragmentation at distances greater than 20 AU.

\section{Initial Conditions}

 Table 1 lists the initial conditions chosen for the five disk models 
presented here. Models 2.0, 1.5, 1.0, 0.5, and 0.1 depict disks around 
protostars with masses of $M_s = 2.0$, 1.5, 1.0, 0.5, and 0.1 $M_\odot$, 
representing future A3, A5, G2, early M, and late M dwarfs, respectively, 
depending on their subsequent accretion of mass. The disk envelopes
are taken to have temperatures ($T_{e}$) between 50 K and 30 K, 
in all cases hotter than the disks themselves, which begin their evolutions 
uniformly isothermal at the initial temperatures ($T_i$) shown in Table 1.
The critical density for differentiating between the disk
and the disk envelope is taken to be $10^{-13}$ g cm$^{-3}$ for models
2.0, 1.5, 1.0, and 0.5, and $10^{-14}$ g cm$^{-3}$ for model 0.1,
which effectively determines the onset of the envelope thermal bath.
Variations in these parameters have been tested by Boss (2007) and
found to have relatively minor effects.
 
 Envelope temperatures of 30 to 50 K appear to be reasonable
bounds for low-mass protostars during quiescent periods 
(Chick \& Cassen 1997). Observations of the DM Tau outer disk, on scales 
of 50 to 60 AU, imply midplane temperatures of 13 to 20 K
(Dartois, Dutrey, \& Guilloteau 2003). Hence the envelope and disk initial
temperatures chosen in Table 1 appear to be reasonable choices for real disks.

 Initially the disks have the density distribution (Boss 1993) of
an adiabatic, self-gravitating, thick disk in near-Keplerian rotation 
about a stellar mass $M_s$

$$ \rho(R,Z)^{\gamma-1} = \rho_o(R)^{\gamma-1} - \biggl( 
{ \gamma - 1 \over \gamma } \biggr) \biggl[
\biggl( { 2 \pi G \sigma(R) \over K } \biggr) Z +
{ G M_s \over K } \biggl( { 1 \over R } - { 1 \over (R^2 + Z^2)^{1/2} }
\biggr ) \biggr], $$

\noindent where $R$ and $Z$ are cylindrical coordinates,
$\rho_o(R)$ is the midplane density, and $\sigma(R)$ is the
surface density. The adiabatic constant is $K = 1.7 \times 10^{17}$ 
(cgs units) and $\gamma = 5/3$ for the initial model; thereafter,
the disk evolves in a nonisothermal manner governed by the energy 
equation and radiative transfer (Boss \& Myhill 1992). The first
adiabatic exponent ($\Gamma_1$) derived from the energy equation
of state for these models varies from 5/3 for temperatures below
100 K to $\sim$ 1.4 for higher temperatures (see Figure 1 in Boss 2007).
The radial variation of the initial midplane density is a power law 
that ensures near-Keplerian rotation throughout the disk

$$\rho_o(R) = \rho_{o4} \biggl( {R_4 \over R} \biggr)^{3/2}, $$

\noindent where $\rho_{o4} = (M_s/M_\odot) \times 10^{-10}$ g cm$^{-3}$ and
$R_4 = 4$ AU. This disk structure is the continuation to 60 AU of the same 
disk used in the $M_s = 1.0 M_\odot$ models of, e.g., Boss (2001, 2003, 2005, 
2006a, 2010). While each disk is initially close to centrifugal balance 
in the radial direction, the use of the Boss (1993) analytical density
distribution, with varied initial disk temperatures, means that the
disks initially contract vertically until a quasi-equilibrium state
is reached (Boss 1998).

 Table 1 lists the resulting disk masses $M_d$ (from 20 AU to 60 AU), 
the disk mass to stellar mass ratios $M_d/M_s$, the initial disk temperatures 
$T_i$, and the initial minimum and maximum values of the Toomre (1964)
$Q$ gravitational stability criterion, increasing monotonically outward from 
unstable $Q = 1.1$ values at 20 AU to marginally stable
$Q \sim 1.6$ at 60 AU. These values of $Q$ were chosen to be low
enough in the inner disk regions to err on the side of clump formation;
higher initial $Q$ values are expected to stifle disk fragmentation.
These models thus represent a first exploration of parameter space
for large-scale disks to establish feasibility. Further work should 
investigate higher $Q$ initial conditions, as disks are expected
to evolve starting from marginally gravitationally unstable ($Q > 1.5$) 
initial conditions (e.g., Boley 2009). Such disks typically also fragment, 
but only after a period of dynamical evolution toward $Q \sim 1$ in 
limited regions, such as dense rings (e.g., Boss 2002).

 Large, massive disks have been detected in regions of low-mass star
formation, such as the 300-AU-scale, $\sim 1 M_\odot$ disk around 
the class O protostar Serpens FIRS 1 (Enoch et al. 2009). Observations
of 11 low- and intermediate-mass pre-main-sequence stars imply that
their circumstellar disks formed with masses in the range from 0.05 $M_\odot$ 
to 0.4 $M_\odot$ (Isella, Carpenter, \& Sargent 2009). These and other 
observations support the choice of the disk masses and sizes assumed 
in the present models.

\section{Results}

 All of the models dynamically evolve in much the same way. Beginning 
from nearly axisymmetric configurations (with initial $m = 1, 2, 3, 4$ 
density perturbations of amplitude 1\%), the disks develop increasingly
stronger spiral arm structures. Eventually these trailing spiral arms 
become distinct enough, through self-gravitational growth and mutual 
collisions, that reasonably well-defined clumps appear and maintain
their identities for some fraction of an orbital period. However,
because the fixed-grid nature of these calculations prevents
the clumps from contracting to much higher densities, the clumps
are doomed to eventual destruction by a combination of thermal
pressure, tidal forces from the protostar, and Keplerian shear.
However, new clumps continue to form and orbit the protostar, suggesting 
that clump formation is inevitable. Previous work (Boss 2005) has shown 
that as the numerical spatial resolution is increased, the survival of
clumps formed by disk instability is enhanced.
While an adaptive-mesh-refinement code would be desirable for
demonstrating that clumps can contract and survive, the present
models, combined with the previous work by Boss (2005), are sufficient
for a first exploration of this region of disk instability parameter
space.

 Figures 1 through 10 show the midplane density and temperature
contours for all five models at a time of $\sim 6 P_{20}$,
where $P_{20}$ is the Keplerian orbital period at the distance
of the inner grid boundary of 20 AU for a protostar with the 
given mass. For models 0.1, 0.5, 1.0. 1.5, and 2.0, respectively,
$P_{20}$ is equal to 283 yr, 126 yr, 89.4 yr, 73.0 yr, and 63.2 yr.
It is clear that clumps have formed by this time in all five models.
However, in order to become a giant planet, clumps must survive 
long enough to contract toward planetary densities. The spherically
symmetric protoplanet models of Helled \& Bodenheimer (2011) suggest 
contraction time scales ranging from $\sim 10^3$ yr to $\sim 10^5$ yr,
depending on the metallicity, for protoplanets with masses from
3 to 7 $M_{Jup}$, so these clumps must survive for many orbital periods
in order to become planets.

 Table 2 lists the estimated properties for those clumps that
appear to be self-gravitating at the earlier time of $\sim 4 P_{20}$,
while Table 3 lists the estimates for the fragments at the time 
of $\sim 6 P_{20}$ depicted in Figures 1 through 10. At both
of these times, clump formation was relatively well-defined, 
so that the clump masses and other properties could be estimated.
Clumps typically first become apparent at $\sim 2 P_{20}$.
Candidate clumps are identified by eye from the equatorial density
contour plots, and then interrogated with a program that allows
the user to select cells adjoining the cell with the maximum density
in order to achieve a candidate clump with an approximately spherical
appearance, in spite of the obvious banana-shape of many clumps.
The fragment masses $M_{frag}$ are then estimated in units of
the Jupiter mass $M_{Jup}$ and compared to the Jeans mass $M_{Jeans}$ 
(e.g., Spitzer 1968) necessary for gravitational stability at the
mean density and temperature of the fragment. Fragments with masses less 
than the Jeans mass are not expected to be stable (Boss 1997).

 The times presented in Figures 1 through 10 range from 387 yr
to 1843 yr, i.e., timescales of order 1000 yr.
Compared to core accretion, where the time scales involved are measured 
typically in millions of yr (e.g., Ida \& Lin 2005) and low mass stars
are in danger of not being able to form gas giant planets at all
(Laughlin, Bodenheimer, \& Adams 2004), the disks around even low
mass stars are able to form clumps on time scales short enough
to permit gas giant protoplanet formation to occur in the
shortest-lived protoplanetary disks.

 Figures 1 through 10 demonstrate that while clump formation
occurs for all of the disks, the clumps that form becoming
increasingly numerous as the mass of the
protostar (and of the corresponding disk) increases, even
though all disks begin their evolution with essentially the
same range of Toomre (1964) $Q$ values. Clearly more massive
disks are able to produce more numerous protoplanets, all
other things being equal. The temperature contour plots show
that significant compressional heating occurs in these
initially isothermal disks as a result of spiral arm 
formation, with the most significant heating occuring near
the edges of the arms and clumps, as more disk gas seeks to
infall onto the spiral structures; the local temperature maxima
do not necessarily fall at the local density maxima. A 
similar effect was found by Boley \& Durisen (2008). This
suggests that clump formation in these relatively cold
outer disks occurs in an opportunistic manner, pulling cold
disk gas together wherever possible. Even the lowest mass
disk in model 0.1 is optically thick, with a vertical optical depth 
of $\sim 5$, while vertical optical depths of $\sim 10^3$
characterize the more massive disks, so some combination of vertical
radiation transport, dynamical motions, and/or convection
(e.g., Boss 2004, Boley \& Durisen 2006, Mayer et al. 2007) 
is necessary for cooling the disk midplane and allowing the 
clumps to continue their contraction toward planetary densities.

 Tables 2 and 3 also list the estimated orbital semimajor axes $a_{frag}$
and eccentricities $e_{frag}$ for the fragments at the same
times as the other fragment properties are estimated. Needless to
say, these values should be taken solely as initial values, as
interactions with the massive disk (e.g., Boss 2005) and the 
other fragments will result in substantial further orbital
evolution. The fragment orbital parameters $a$ and $e$ were calculated 
using each fragment's radial distance $r$, average radial velocity $v_r$, 
and average azimuthal velocity $v_\phi$ (both derived from the
total momentum of the clump), along with the model's 
stellar mass $M_s$, and inserting these values into these 
equations for a body on a Keplerian orbit (Danby 1988):

$$ a = {G M_s \over 2} \biggl( {G M_s \over r} - {v_r^2 \over 2} - 
{v_\phi^2 \over 2} \biggr)^{-1}, $$

$$ e = \biggl( 1 - {r^2 v_\phi^2 \over G M_s a} \biggr)^{1/2}, $$

\noindent
where $G$ is the gravitational constant.

 Figures 11 and 12 depict the midplane density and temperature
profiles for two of the clumps that form in model 1.0, as shown
in Figures 5 and 6. The fragment in Figure 11 is less well-defined
than that in Figure 12, yet still has an estimated mass of 3.8 $M_{Jup}$,
well above its relevant Jeans mass of 1.8 $M_{Jup}$. The fragment in 
Figure 12 has an estimated mass of 2.5 $M_{Jup}$, also 
well above its relevant Jeans mass of 2.2 $M_{Jup}$. These figures
show that the higher density fragment in Figure 12 has resulted
in a higher temperature interior, while the lower density fragment
in Figure 11 has not yet reached similar internal temperatures,
though in both cases the maximum fragment temperatures occur
close to their edges. Similar plots characterize all of the 
fragments found in these models.

 Figures 13, 14, 15, and 16 plot the resulting estimates of the
initial protoplanet masses, semimajor axes, and eccentricities
as a function of protostar or protoplanetary disk mass, 
for the fragments in Tables 2 and 3
where the fragment mass is equal to or greater than the Jeans
mass. Given the uncertain future evolution of the fragments
as they attempt to survive and become true protoplanets, these
values should be taken only as reasonable estimates based on
the present set of models, subject to the inherent assumptions
about the initial disk properties. Boley et al. (2010), for
example, found that clumps on highly eccentric orbits could
be tidally disrupted at periastron. In addition, surviving
fragments are likely to gain substantially more disk gas
mass during their orbital evolution (``type IV non-migration'')
in a marginally gravitationally unstable disk (Boss 2005).
Nevertheless, Figure 13 makes it clear that
disk instability is capable of leading to gas giant protoplanet
formation around protostars with masses in the range from 0.1
to 2.0 $M_\odot$, with more protoplanets forming as the mass
of the protostar and its disk increases: perhaps only a
single protoplanet for a 0.1 $M_\odot$ protostar, but as
many as six for a 2.0 $M_\odot$ protostar. The masses of 
the protoplanets appear to increase with the stellar and disk
mass (Figures 13 and 14); the typical initial protoplanet mass 
increased from $\sim 1 M_{Jup}$ to $\sim 3 M_{Jup}$ over the 
range of models 0.1 to 2.0. 

 Given the large disk masses, it is likely that the final 
protoplanet masses will similarly increase with time,
as those protoplanets accrete mass from a massive reservoir 
of gas and dust. This growth will be limited though by the
angular momentum of the disk gas that the protoplanet
is trying to accrete (Boley et al. 2010).
Nevertheless, if one wishes to use these models to explain
the formation of giant planets with minimum masses similar
to those estimated for HR 8799, i.e., 5 to 7 $M_{Jup}$, 
then the outer disk gas must be removed prior to growth
of the protoplanets to unacceptably large masses.
Photoevaporation of the outer disk by FUV and EUV fluxes
from nearby massive (OB) stars is a likely means for achieving
this timely disk gas removal, on a time scale of
$\sim 10^5$ yr (e.g., Balog et al. 2008; Mann \& Williams 2009, 2010).
If the A5V star HR 8799 formed in a region of high mass star formation,
as in the case for the majority of stars, the outer disk gas
should disappear within $\sim 10^5$ yr.
If giant protoplanets cannot accrete mass from the disk
at a rate higher than $\sim 10^{-4} M_{Jup}$ yr$^{-1}$, as
argued by Nelson \& Benz (2003), then the maximum amount of
disk gas that could be accreted in $10^5$ yr or less would
be $\sim 10 M_{Jup}$. A mass gain no greater than this appears
to be roughly consistent with the range of masses estimated
for the four planets in HR 8799 (Marois et al. 2008, 2010),
which could be as high as 13 $M_{Jup}$.

 Figures 15 and 16 show that these protoplanets begin their
existence with orbital semimajor axes in the range of $\sim$ 30 AU
to $\sim$ 70 AU and orbital eccentricities from $\sim$ 0 to $\sim$ 0.35.
Only upper limits of $\sim 0.4$ exist for the orbital eccentricities of
the HR 8799 system (Figure 16), comfortably above the model estimates.
The initial orbital eccentricities appear to vary slightly with stellar mass,
with eccentricities dropping as the stellar mass increases, though
this hint is largely due to the higher eccentricities found in
model 0.1. The semimajor axes show a similar slight trend of 
decreasing with stellar mass, though both of these effects may
be more of a result of small number statistics than of any
robust physical mechanism at work.

 Just as there is a danger that protoplanets formed in a massive
disk could grow to become brown dwarfs, unless prevented from
doing so by removal of the outer disk gas through photoevaporation,
there is a danger that the protoplanets might suffer inward orbital
migration due to interactions with the disk gas prior to its
removal. However, models of the interactions of protoplanets
with marginally gravitationally unstable disks (Boss 2005) have
shown that the protoplanets experience an orbital evolution
that is closer to a random walk (type IV non-migration) than
to the classic monotonically inward (or outward) evolution due
to type II migration, where the planet clears a gap in the disk and
then must move in the same direction as the surrounding disk gas.
Hence, the outer disk protoplanets formed in these models
need not be expected to suffer major inward or outward migration
prior to photoevaporation of the outer disk, though clearly 
this possibility is deserving of further study.

 The close-packing in semimajor axis of the fragments in model 2.0 
(Figures 9 and 15) makes it clear that these protoplanets
will interact gravitationally with each other (as well as with
the much more massive disk), resulting in mutual close encounters
and scattering of protoplanets to orbits with larger and smaller
semimajor axes than their initial values (Figure 15). The
evolution during this subsequent phase is best described with
a fixed-grid code by using the virtual protoplanet technique,
where the fragments are replaced by point mass objects that
orbit and interact with the disk and each other (Boss 2005).
Models that continue the present models with the virtual
protoplanet technique are now underway and will be presented
in a future paper.

\section{Discussion}

 Most disk instability models have focused on forming giant planets 
similar to those in our Solar System, and hence have studied disks 
with outer radii of 20 AU (e.g., Boss 2001; Mayer et al. 2007). 
Boss (2003) found that disk instability could lead to the formation
of self-gravitating clumps with initial orbital semimajor axes
of $\sim 20$ AU in disks with outer radii of 30 AU. On the
other hand, Boss (2006a) found no strong tendency for clump formation
in disks extending from 100 AU to 200 AU. In both cases these
models assumed 1 $M_\odot$ central protostars. Model 1.0 in the
present work shows that when the disk is assumed to extend from
20 AU to 60 AU, clumps are again expected to be able to form,
with initial semimajor axes of $\sim$ 30 AU to $\sim$ 45 AU
(Figure 15). Taken together, these models imply that for a
1 $M_\odot$ protostar at least, disk instability might be able
to form gaseous protoplanets with initial semimajor axes anywhere
inside $\sim$ 50 AU. When multiple protoplanets form, as is
likely to be the case for stars more massive than M dwarfs,
subsequent gravitational interactions are likely to result in
at least a few protoplanets being kicked out to orbits with
semimajor axes greater than 50 AU. 

 Other authors have also considered the evolution of gravitationally
unstable disks with outer radii much greater than 20 AU.
Stamatellos \& Whitworth (2009a,b) used a smoothed particle
hydrodynamics (SPH) code with radiative transfer in the diffusion 
approximation to model the evolution of disk instabilities in disks 
with the same mass as the central protostar: $M_d = M_s = 0.7 M_\odot$. 
The disks extended from 40 AU to 400 AU, with initial Toomre $Q$ 
values of 0.9 throughout, making them initially highly gravitationally
unstable. As expected, these disks rapidly fragmented into multiple
clumps, which often grew to brown dwarf masses (i.e., greater than
13 $M_{Jup}$) or higher, with final orbital radii as large as 800 AU. 
Their results are in general agreement with the present results, 
though the major differences in the initial disk assumptions 
preclude a detailed comparison. 

 Boley et al. (2010) used an SPH code to demonstrate multiple 
fragment formation at distances from $\sim$ 50 AU to $\sim$ 100 AU 
from a 0.3 $M_\odot$ star in a disk with a radius of 510 AU and 
a mass of 0.19 $M_\odot$. Given the large disk mass to stellar
mass ratio of 0.63, the formation of several clumps with initial masses
of 3.3 $M_{Jup}$ and 1.7 $M_{Jup}$ is basically consistent with the
present results for model 0.5. 

 The result that clump formation depends on protostellar mass,
with models 0.1 and 0.5 forming fewer clumps than models 1.0,
1.5, and 2.0, is consistent with the results presented by
Boss (2006b), who studied the evolution of disks with outer
radii of 20 AU around protostars with masses of 0.1 $M_\odot$
and 0.5 $M_\odot$. Boss (2006b) found that clumps could form for
both protostar masses, but that while several clumps formed
for the 0.5 $M_\odot$ protostar, typically only a single
clump formed for the 0.1 $M_\odot$ protostar, similar to the
results in models 0.1 and 0.5 for much larger radii disks. 
Thus, while not zero, the chances for giant planet formation
by disk instability appear to decrease with stellar mass
in the range of 0.5 $M_\odot$ to 0.1 $M_\odot$. A simple
explanation for this outcome may be that given the assumption
of disk masses that scale with protostellar masses, the number
of Jupiter-mass protoplanets that could form by disk instability
increases with the number of Jupiter-masses of disk gas available 
for their formation: e.g., the disk mass for model 2.0 is taken to
be 7.5 times that of model 0.1.

 Nero \& Bjorkman (2009) used analytical models to study
fragmentation in suitably massive protoplanetary disks,
finding that their estimated cooling times were over an
order of magnitude shorter than those estimates previously
by Rafikov (2005), a result consistent with that of Boss (2005).
Nero \& Bjorkman (2009) found that the outermost planet
around HR 8799 was likely to have formed by a disk instability,
but that the two closer-in planets were not, a conclusion
at odds with the results of the present numerical calculations.
The different outcomes appear to be a result of different
assumptions about the initial disk density and temperature
profiles, dust grain opacities, and use of a cooling time
argument rather than detailed radiative transfer and
hydrodynamics. Recently the use of cooling times to
depict the thermodynamics of protoplanetary disks has
been called in question by the three dimensional hydrodynamical
models of Meru \& Bate (2010, 2011), who found that previous
calculations relied on an overly simplistic cooling time
argument, and that when sufficiently high spatial resolution
was employed, even disks previously thought to be stable 
underwent fragmentation into clumps.

 Finally, it is interesting to note an observational prediction.
Helled \& Bodenheimer (2010) have modeled the capture of solids
by gas giant protoplanets formed at distances similar to those
of the four planet candidates in HR 8799 (Marois et al. 2008, 2010).
They found that because such massive protoplanets contract rapidly
on time scales of only $\sim 10^4$ yr, few planetesimals can be
captured by gas drag in their outer envelopes, leading to a
prediction that the bulk compositions of these four objects
should be similar to that of their host stars if they formed by
disk instability. HR 8799 has a low metallicity ([M/H] = -0.47;
Gray \& Kaye 1999), so the HR 8799 objects are expected to
be similarly metal-poor, unless the protoplanets are able to
form in a dust-rich region of the disk (Boley \& Durisen 2010).

\section{Conclusions}

 The present set of models has shown that disk instability is capable
of the rapid formation of giant planets on relatively wide orbits around 
protostars with masses in the range from 0.1 $M_\odot$ to 2.0 $M_\odot$.
While the number of protoplanets formed by disk instability 
appears to increase with the mass of the star (and hence of the
assumed protoplanetary disk), even late M dwarf stars might be
able to form gas giants on wide orbits, provided that suitably
gravitationally unstable disks exist in orbit around them.
These results suggest that direct imaging searches for gas giant
planets on wide orbits around low mass stars are likely to continue 
to bear fruit; the protoplanet candidates detected to date do not 
appear to be rare oddballs unexplainable by theoretical models
of planetary system formation. HR 8799's four planets in particular
appear to be broadly consistent with formation by disk instability,
though clearly further study of the formation of this key 
planetary system is warranted.

 I thank the two referees and the scientific editor, Eric 
Feigelson, for valuable improvements to both the manuscript
itself, and, more importantly, to the choice of initial conditions 
for the models, Sandy Keiser for computer systems support, and John 
Chambers for advice on orbit determinations. This research was 
supported in part by NASA Planetary Geology and Geophysics grant 
NNX07AP46G, and is contributed in part to NASA Astrobiology Institute
grant NNA09DA81A. The calculations were performed on the flash
cluster at DTM.

\clearpage
\begin{deluxetable}{lccccccc}
\tablecaption{Initial conditions for the models.}
\label{tbl-1}
\tablewidth{0pt}
\tablehead{\colhead{Model}
& \colhead{$M_s/M_\odot$}
& \colhead{$M_d/M_\odot$}
& \colhead{$M_d/M_s$}
& \colhead{$T_i$} 
& \colhead{$T_e$} 
& \colhead{$Q_{min}$}
& \colhead{$Q_{max}$}}
\startdata

2.0 &  2.0 &  0.21  &   0.11 &   40. &  50. & 1.13 & 1.71  \\    

1.5 &  1.5 &  0.17  &   0.11 &   35. &  40. & 1.12 & 1.67  \\    

1.0 &  1.0 &  0.13  &   0.13 &   30. &  30. & 1.13 & 1.68 \\    

0.5 &  0.5 &  0.083 &   0.17 &   22. &  30. & 1.12 & 1.61 \\ 

0.1 &  0.1 &  0.028 &   0.28 &   11. &  30. & 1.11 & 1.47 \\     
   
\enddata
\end{deluxetable}

\clearpage

\begin{deluxetable}{lcccc}
\tablecaption{Fragment properties at time $\sim 4 P_{20}$.}
\label{tbl-2}
\tablewidth{0pt}
\tablehead{\colhead{Model}
& \colhead{$M_{disk}/M_\odot$}
& \colhead{$M_{frag}/M_{Jup}$}
& \colhead{$a_{frag}/$AU}
& \colhead{$e_{frag}$}}
\startdata

2.0 & 0.21 & 4.6 & 33.8 & 0.107 \\
2.0 & 0.21 & 2.7 & 36.7 & 0.104 \\
2.0 & 0.21 & 3.5 & 37.6 & 0.106 \\
2.0 & 0.21 & 3.5 & 33.5 & 0.122 \\
2.0 & 0.21 & 4.0 & 36.4 & 0.135 \\
2.0 & 0.21 & 3.4 & 35.6 & 0.149 \\
1.5 & 0.17 & 5.1 & 35.2 & 0.191 \\
1.5 & 0.17 & 4.2 & 31.2 & 0.151 \\
1.5 & 0.17 & 1.6 & 32.1 & 0.077 \\
1.5 & 0.17 & 1.8 & 31.8 & 0.070 \\
1.0 & 0.13 & 2.8 & 35.1 & 0.166 \\
1.0 & 0.13 & 1.8 & 32.0 & 0.119 \\
1.0 & 0.13 & 2.9 & 40.5 & 0.117 \\
0.5 & .083 & 1.4 & 43.2 & 0.174 \\
0.1 & .028 & .74 & 48.7 & 0.350 \\

\enddata
\end{deluxetable}

\clearpage

\begin{deluxetable}{lcccc}
\tablecaption{Fragment properties at time $\sim 6 P_{20}$.}
\label{tbl-3}
\tablewidth{0pt}
\tablehead{\colhead{Model}
& \colhead{$M_{disk}/M_\odot$}
& \colhead{$M_{frag}/M_{Jup}$}
& \colhead{$a_{frag}/$AU}
& \colhead{$e_{frag}$}}
\startdata

2.0 & 0.21 & 1.4 & 43. & 0.019 \\ 
2.0 & 0.21 & 3.1 & 39. & 0.051 \\ 
2.0 & 0.21 & 2.9 & 43. & 0.043 \\
1.5 & 0.17 & 4.2 & 39. & 0.19  \\
1.5 & 0.17 & 3.0 & 48. & 0.22  \\
1.5 & 0.17 & 2.4 & 41. & 0.047 \\
1.5 & 0.17 & 4.9 & 67. & 0.22  \\
1.0 & 0.13 & 3.0 & 39. & 0.12  \\
1.0 & 0.13 & 2.0 & 37. & 0.031 \\
1.0 & 0.13 & 3.8 & 44. & 0.11  \\
1.0 & 0.13 & 2.5 & 45. & 0.14  \\
0.5 & .083 & 2.1 & 44. & 0.092 \\
0.5 & .083 & 1.9 & 51. & 0.24  \\
0.1 & .028 & .91 & 45. & 0.33  \\
0.1 & .028 & .80 & 42. & 0.33  \\

\enddata
\end{deluxetable}

\clearpage

\begin{figure}
\vspace{-2.0in}
\plotone{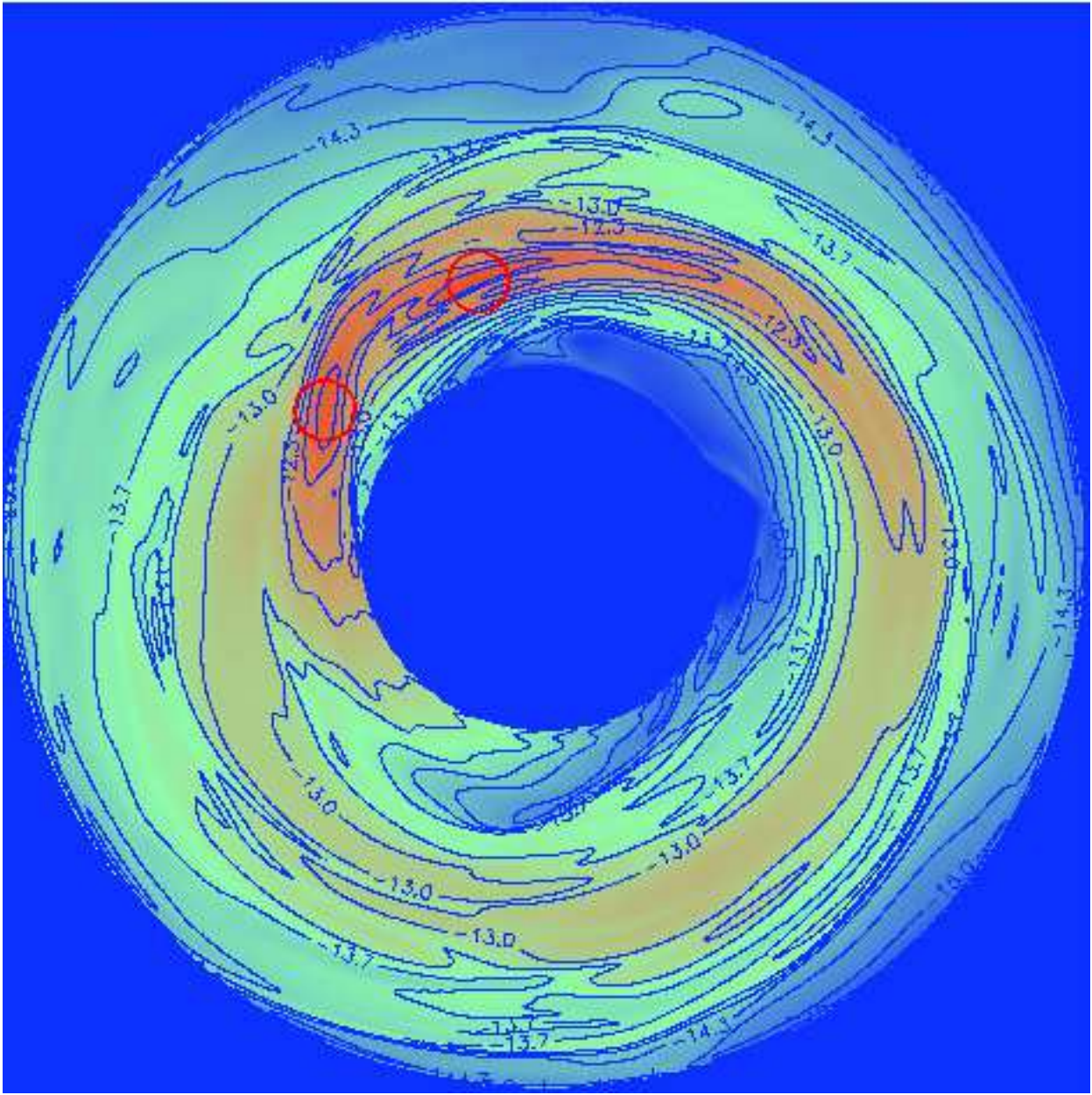}
\caption{Equatorial density contours for model 0.1 after 1843 yr of 
evolution. In this plot and subsequent density contour plots, each contour 
represents a change in density by factor of about 2. The disk has an inner
radius of 20 AU and an outer radius of 60 AU. Red circles denote the
fragments listed in Table 3.}
\end{figure}

\begin{figure}
\vspace{-2.0in}
\plotone{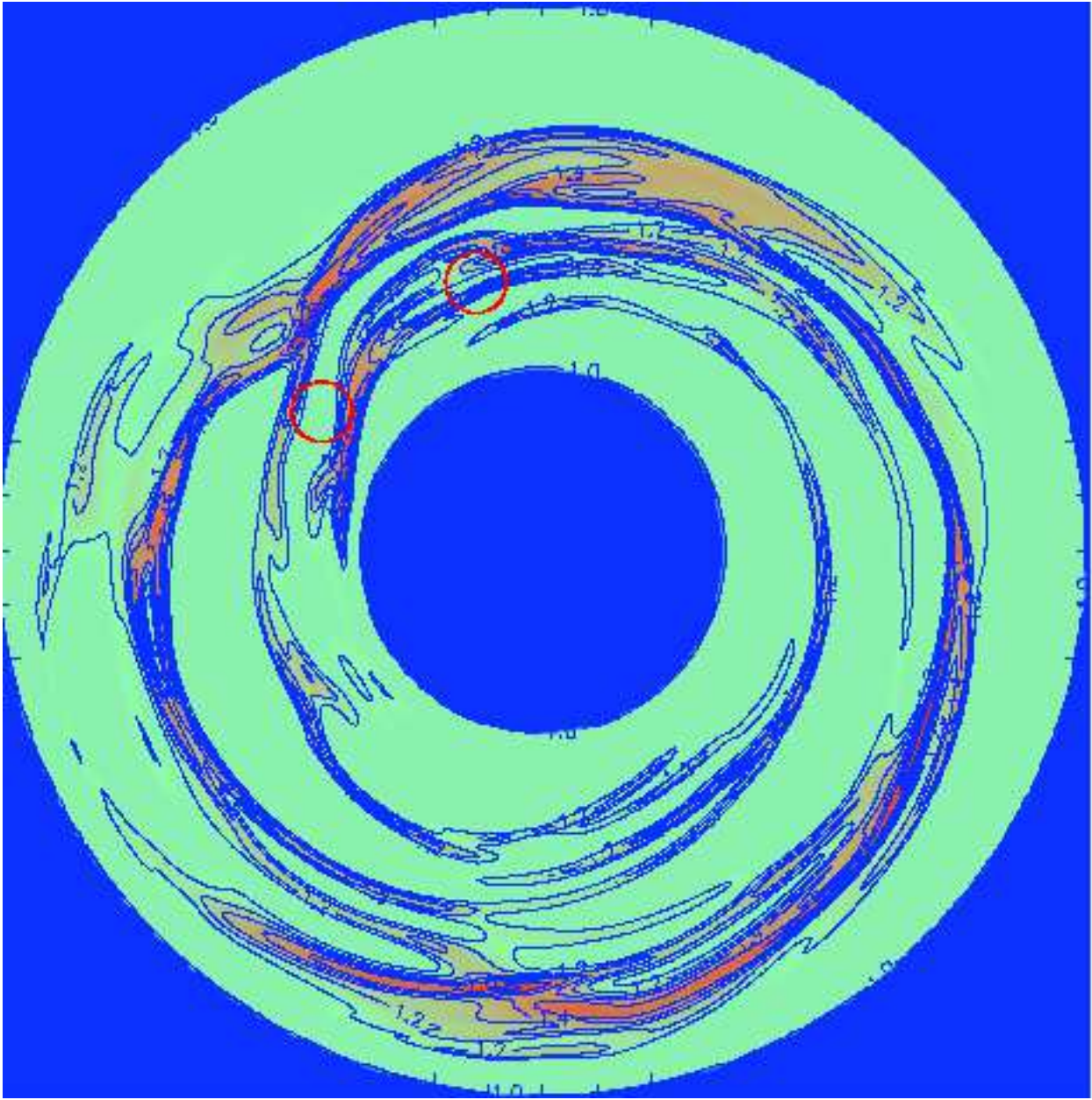}
\caption{Equatorial temperature contours for model 0.1 after 1843 yr of 
evolution. In this plot and subsequent temperature contour plots, each contour
represents a change in temperature by a factor of about 1.3.}
\end{figure}

\clearpage

\begin{figure}
\vspace{-2.0in}
\plotone{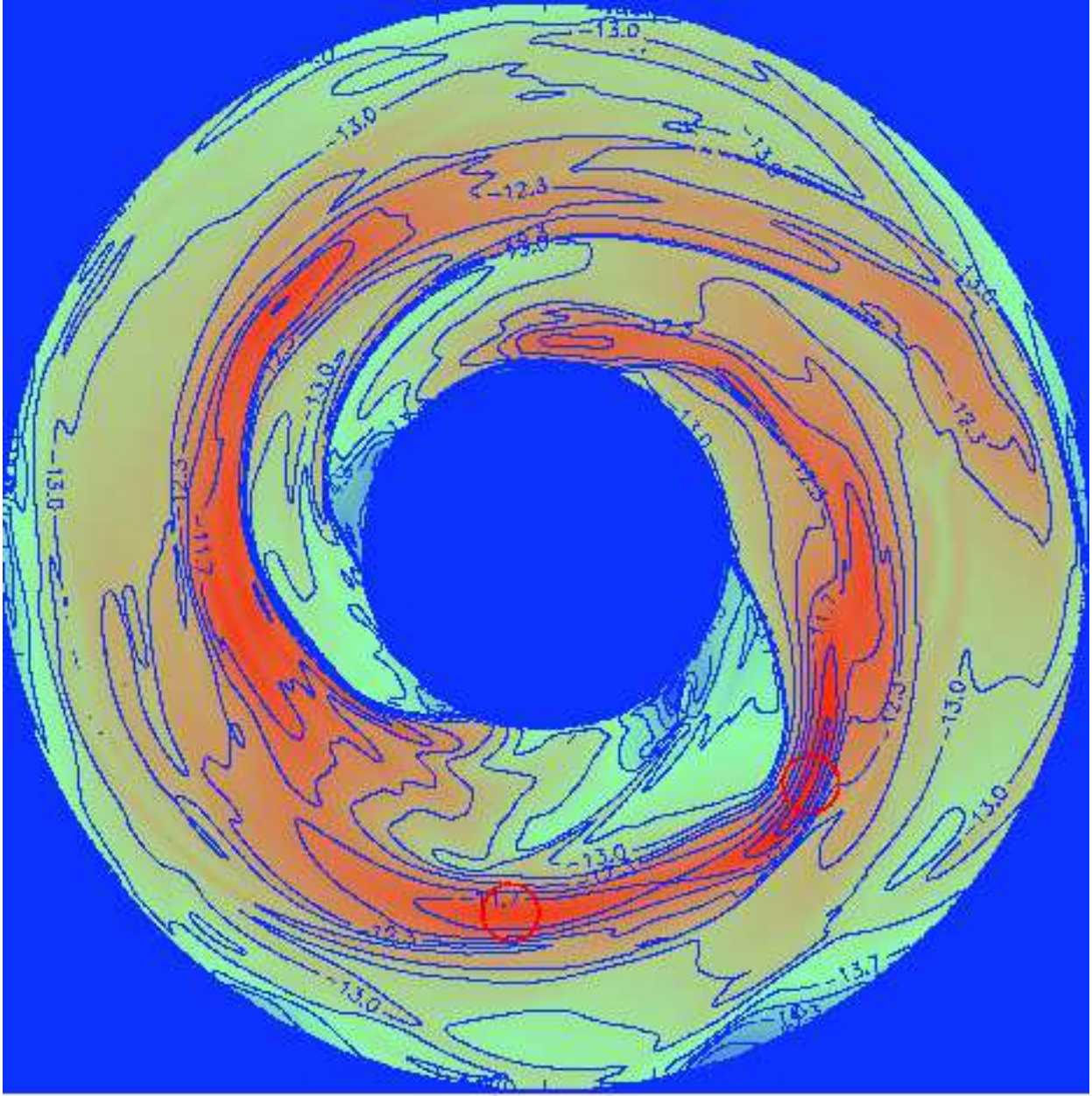}
\caption{Equatorial density contours for model 0.5 after 771 yr of 
evolution.}
\end{figure}
  
\begin{figure}
\vspace{-2.0in}
\plotone{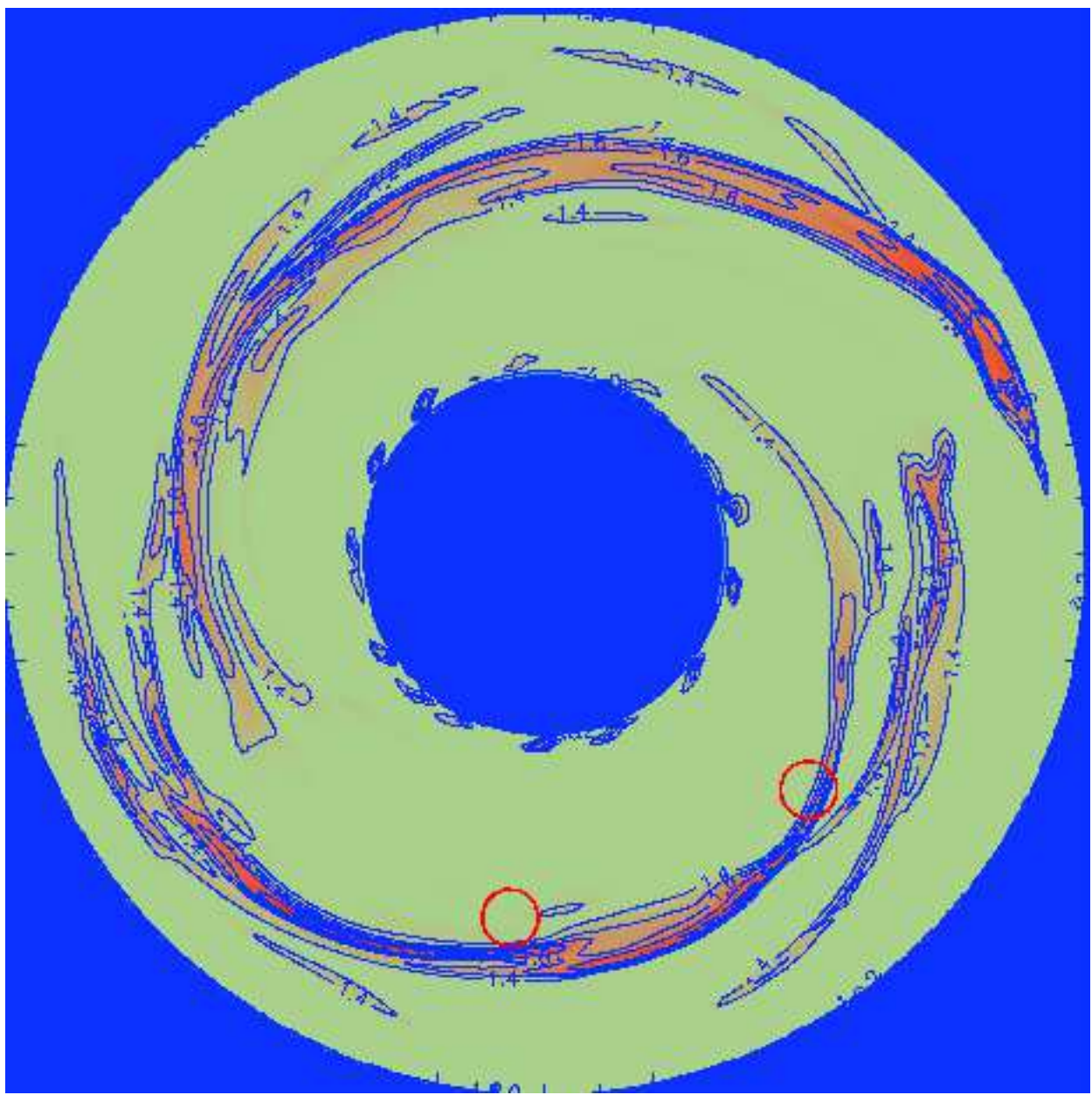}
\caption{Equatorial temperature contours for model 0.5 after 771 yr of 
evolution.}
\end{figure}

\clearpage

\begin{figure}
\vspace{-2.0in}
\plotone{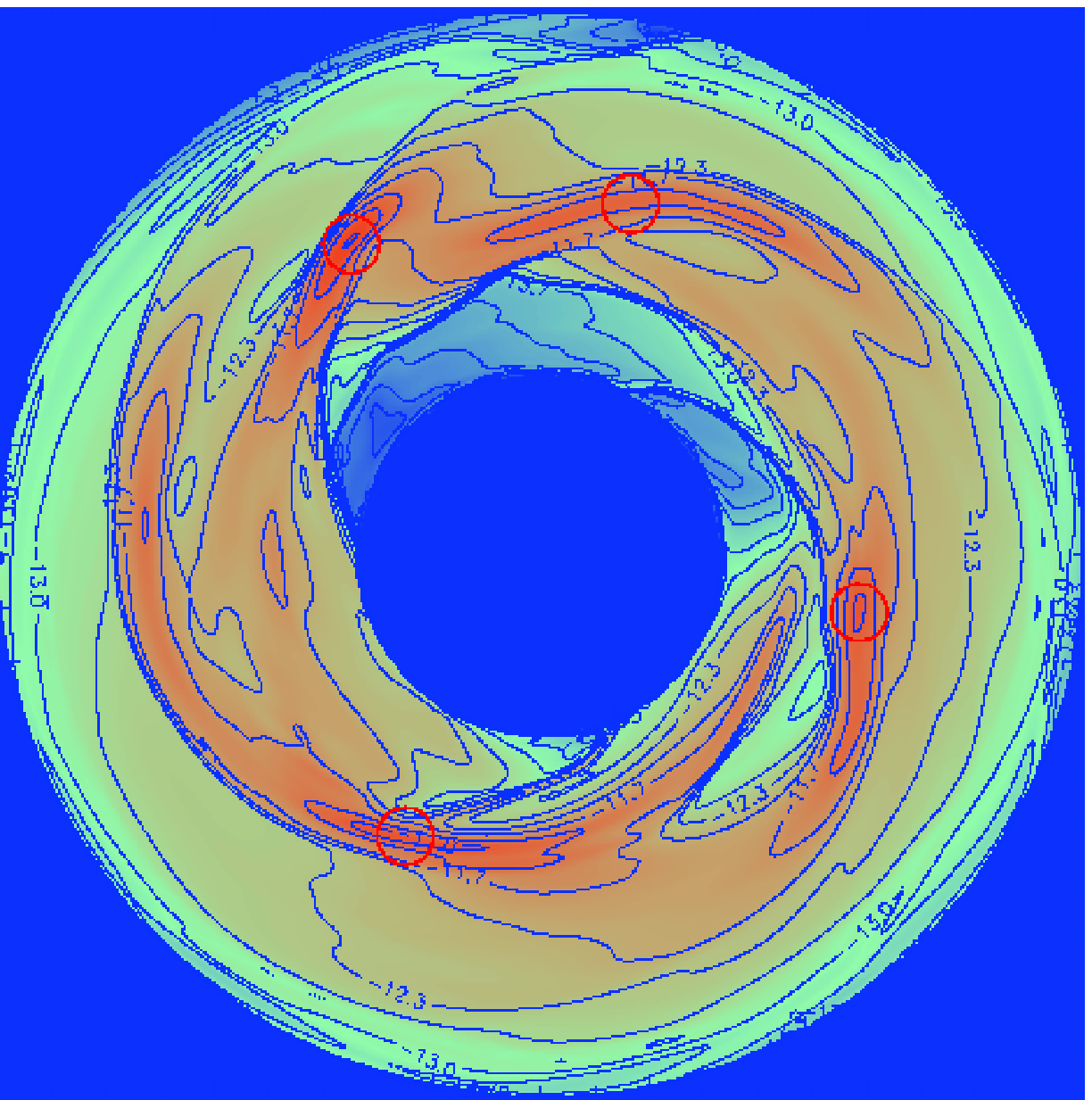}
\caption{Equatorial density contours for model 1.0 after 547 yr of 
evolution.}
\end{figure}

\begin{figure}
\vspace{-2.0in}
\plotone{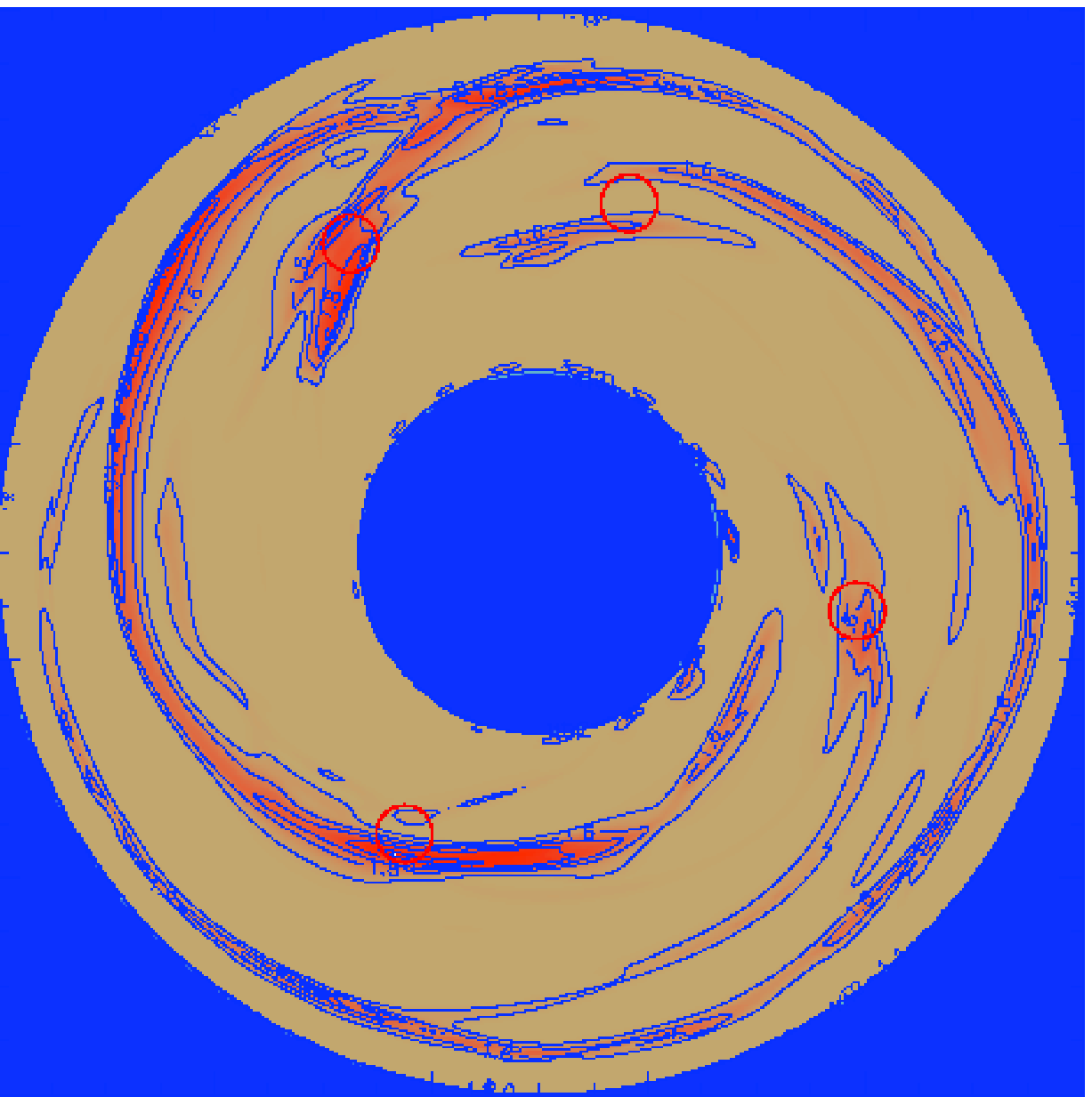}
\caption{Equatorial temperature contours for model 1.0 after 547 yr of 
evolution.}
\end{figure}

\clearpage

\begin{figure}
\vspace{-2.0in}
\plotone{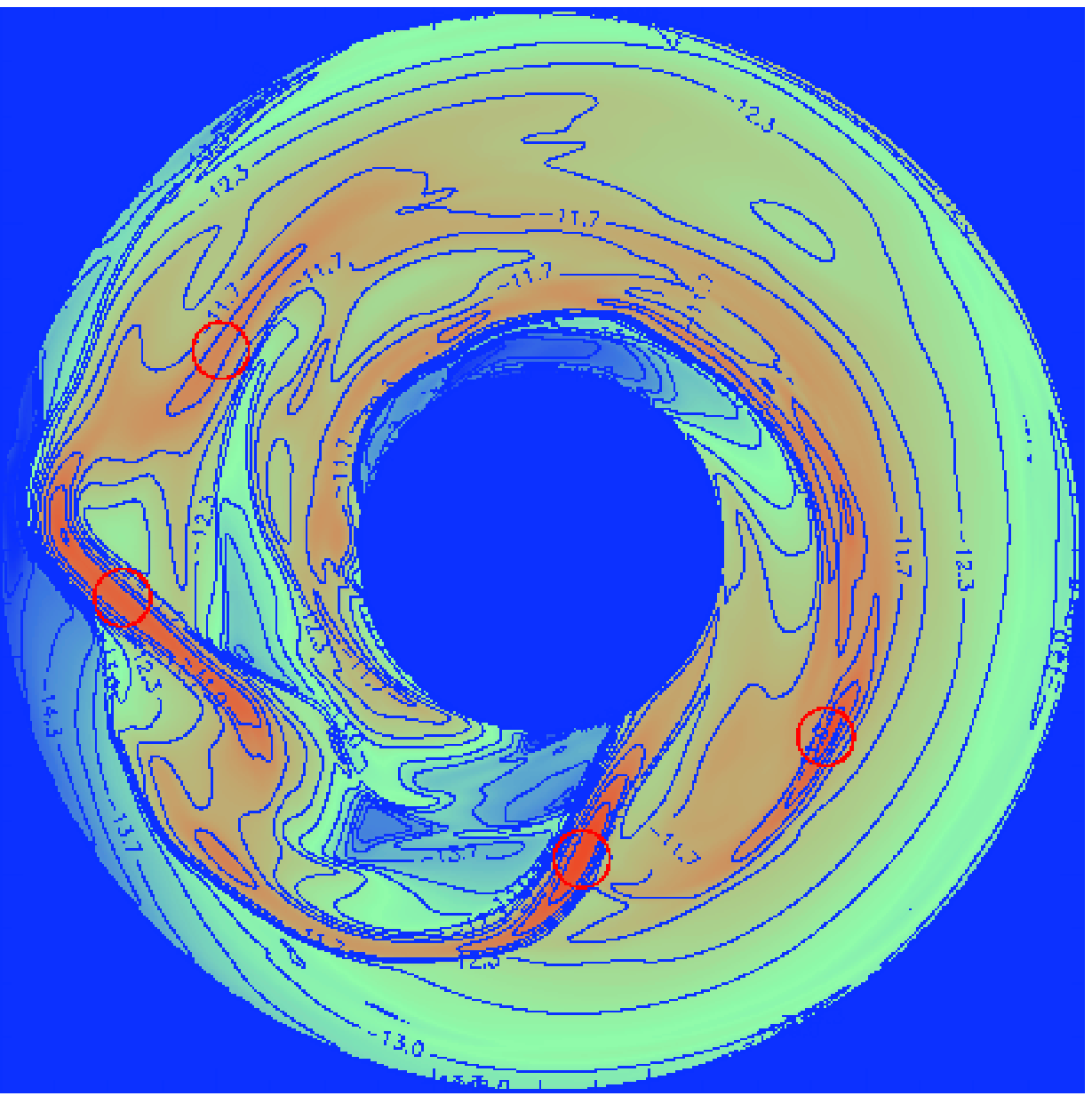}
\caption{Equatorial density contours for model 1.5 after 446 yr of 
evolution.}
\end{figure}

\begin{figure}
\vspace{-2.0in}
\plotone{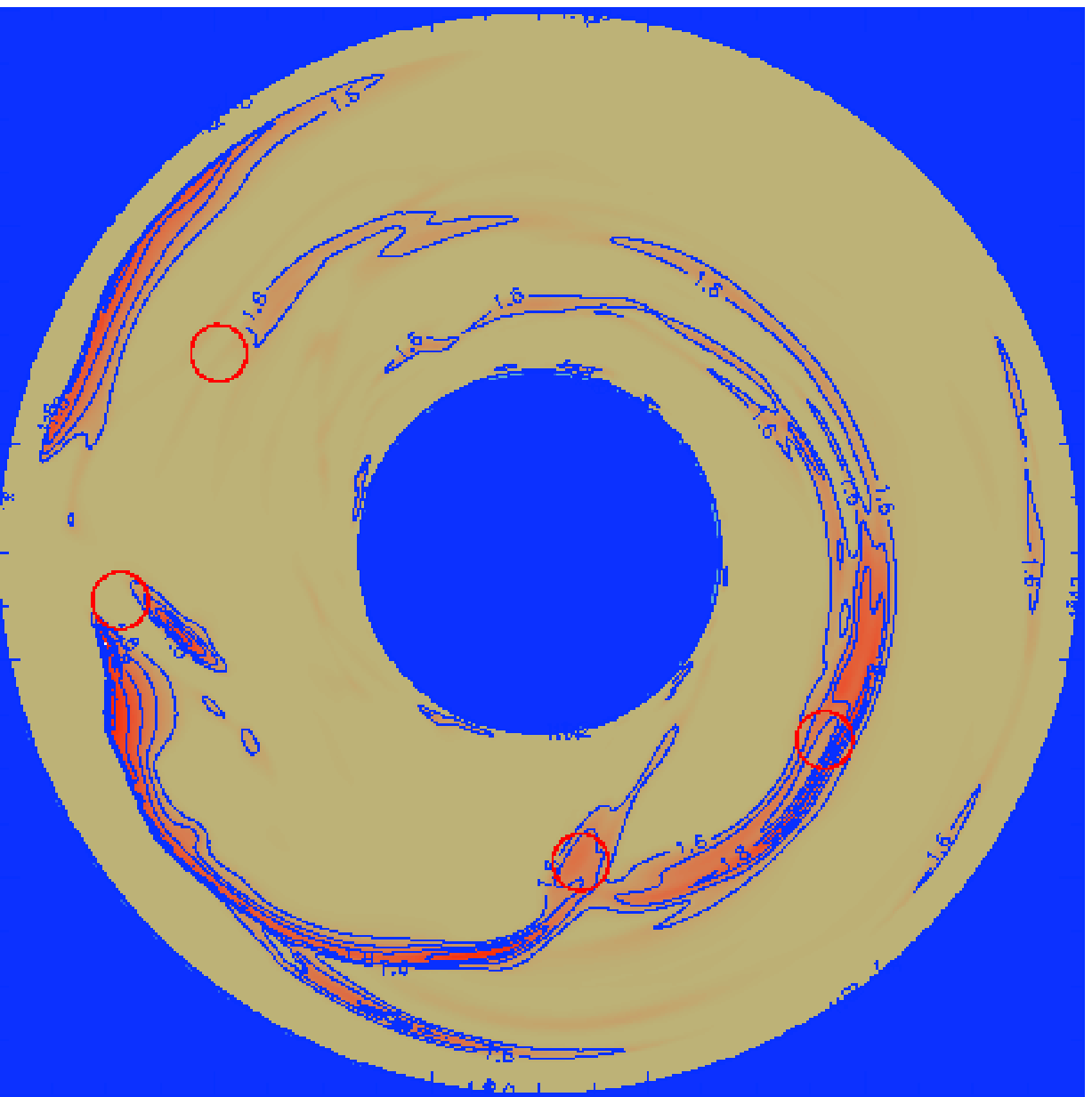}
\caption{Equatorial temperature contours for model 1.5 after 446 yr of 
evolution.}
\end{figure}

\clearpage

\begin{figure}
\vspace{-2.0in}
\plotone{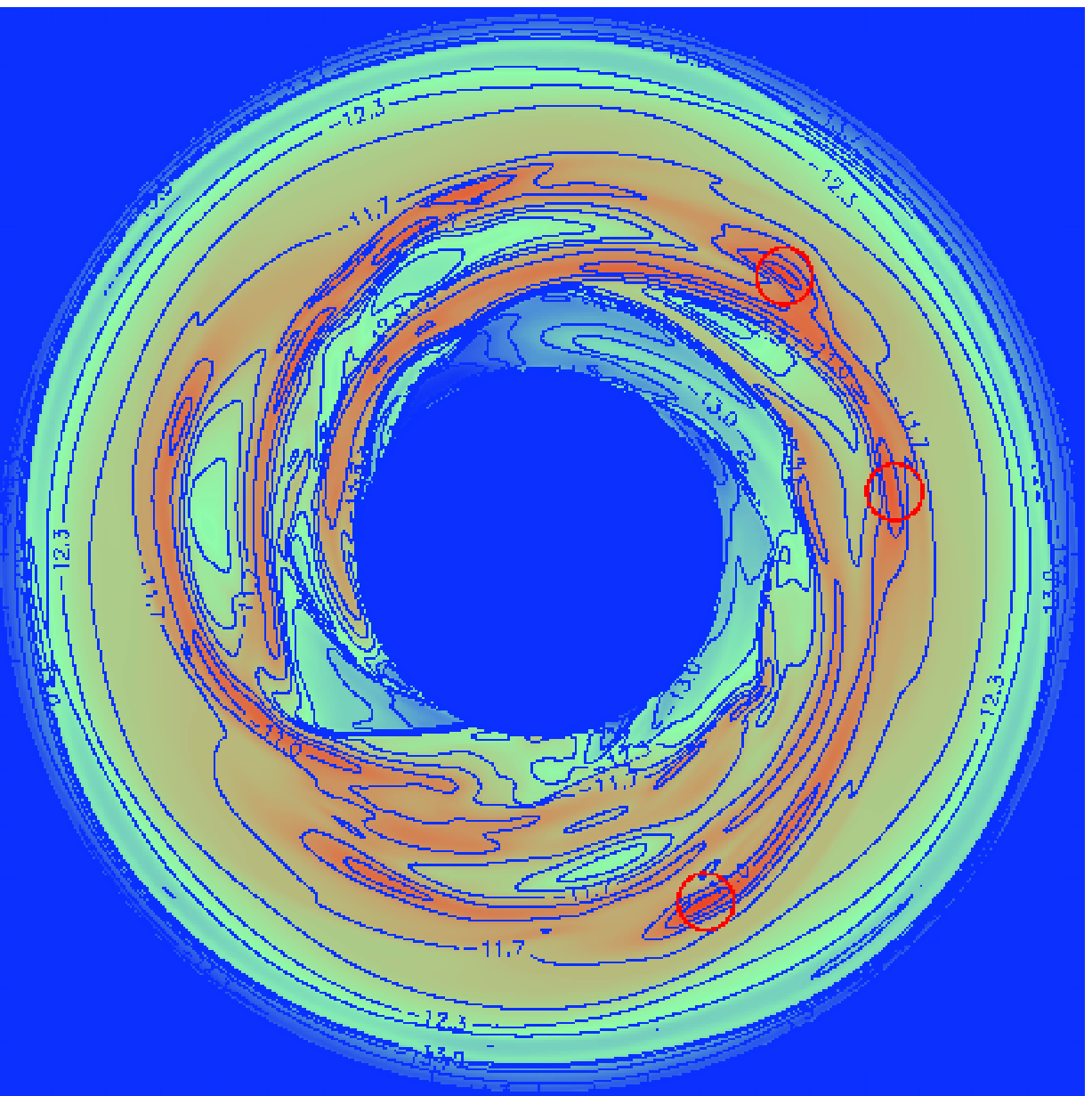}
\caption{Equatorial density contours for model 2.0 after 387 yr of 
evolution.}
\end{figure}

\begin{figure}
\vspace{-2.0in}
\plotone{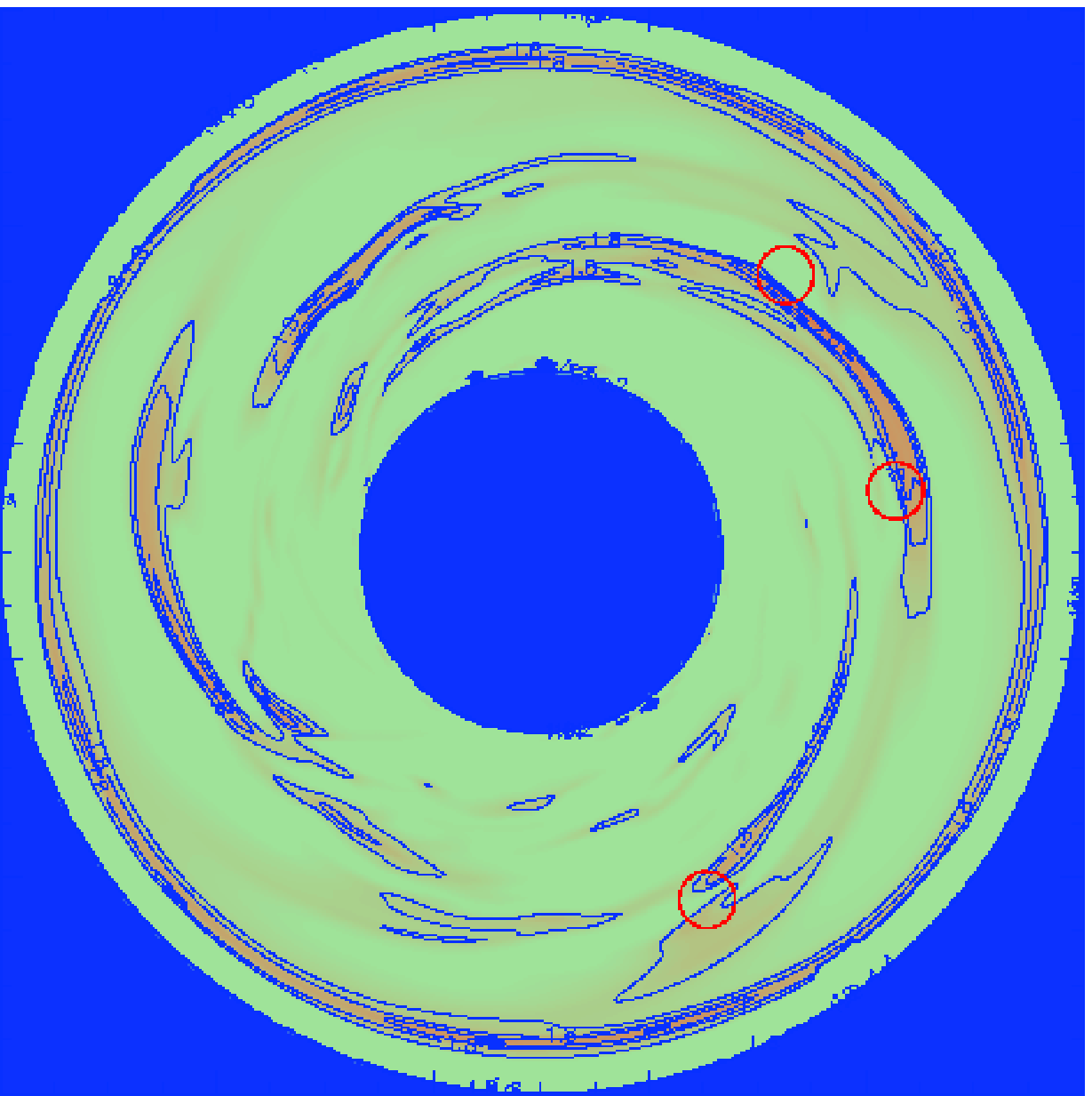}
\caption{Equatorial temperature contours for model 2.0 after 387 yr of 
evolution.}
\end{figure}

\clearpage

\begin{figure}
\vspace{-2.0in}
\plotone{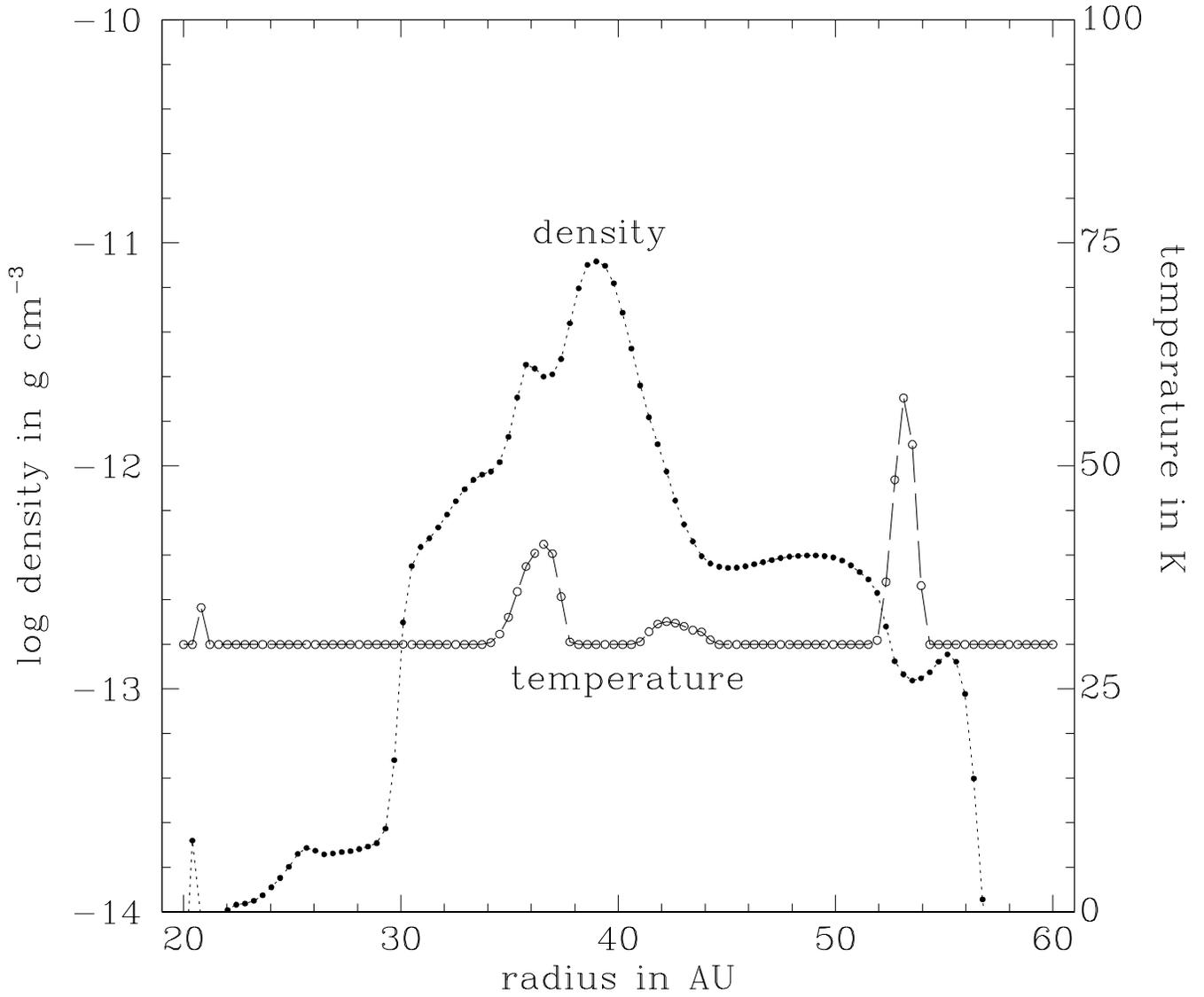}
\caption{Midplane density and temperature radial profiles for the clump at
12 midnight in Figures 5 and 6 for model 1.0.}
\end{figure}

\clearpage

\begin{figure}
\vspace{-2.0in}
\plotone{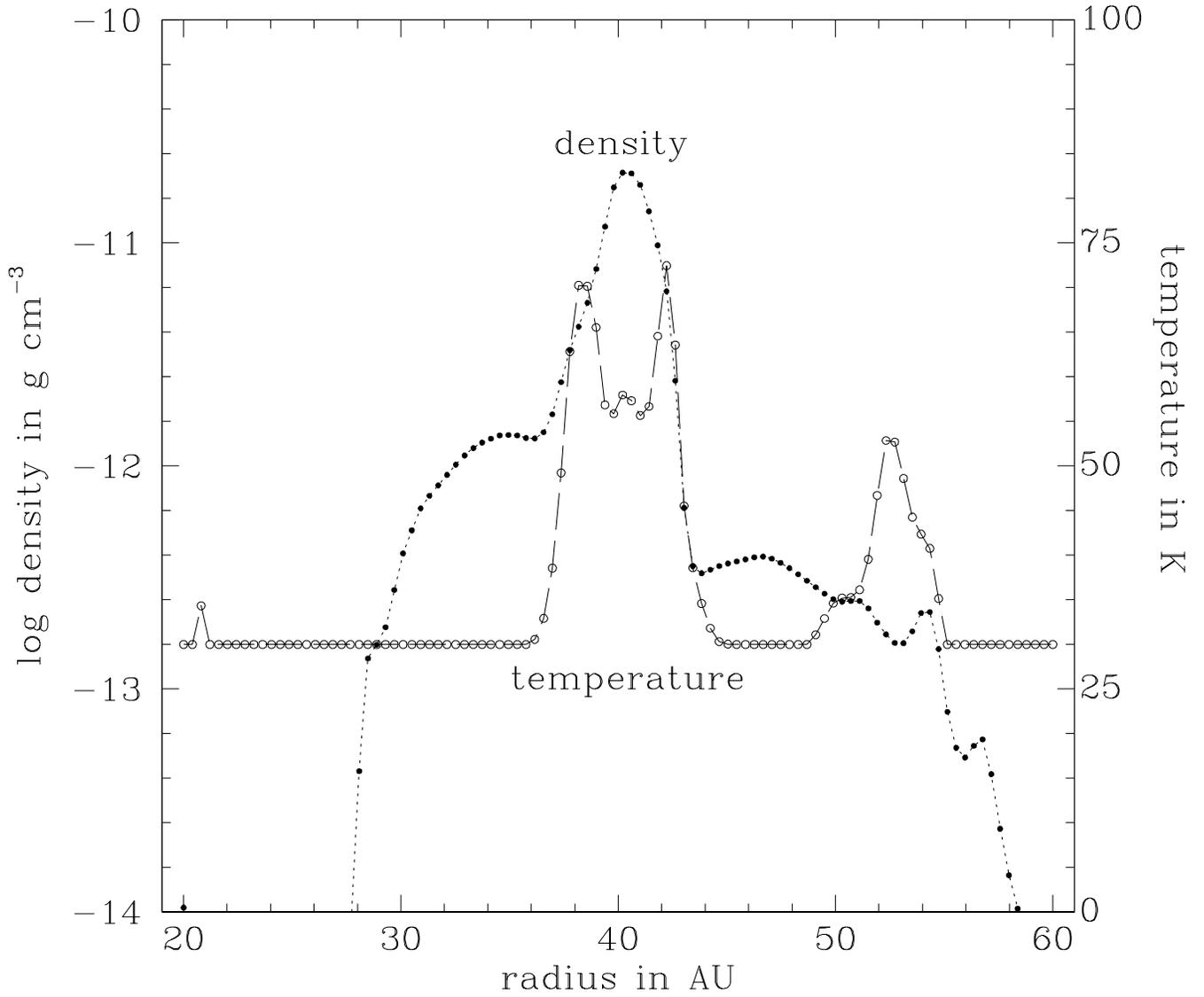}
\caption{Midplane density and temperature radial profiles for the clump at
11 o'clock in Figures 5 and 6 for model 1.0.}
\end{figure}

\clearpage

\begin{figure}
\vspace{-2.0in}
\plotone{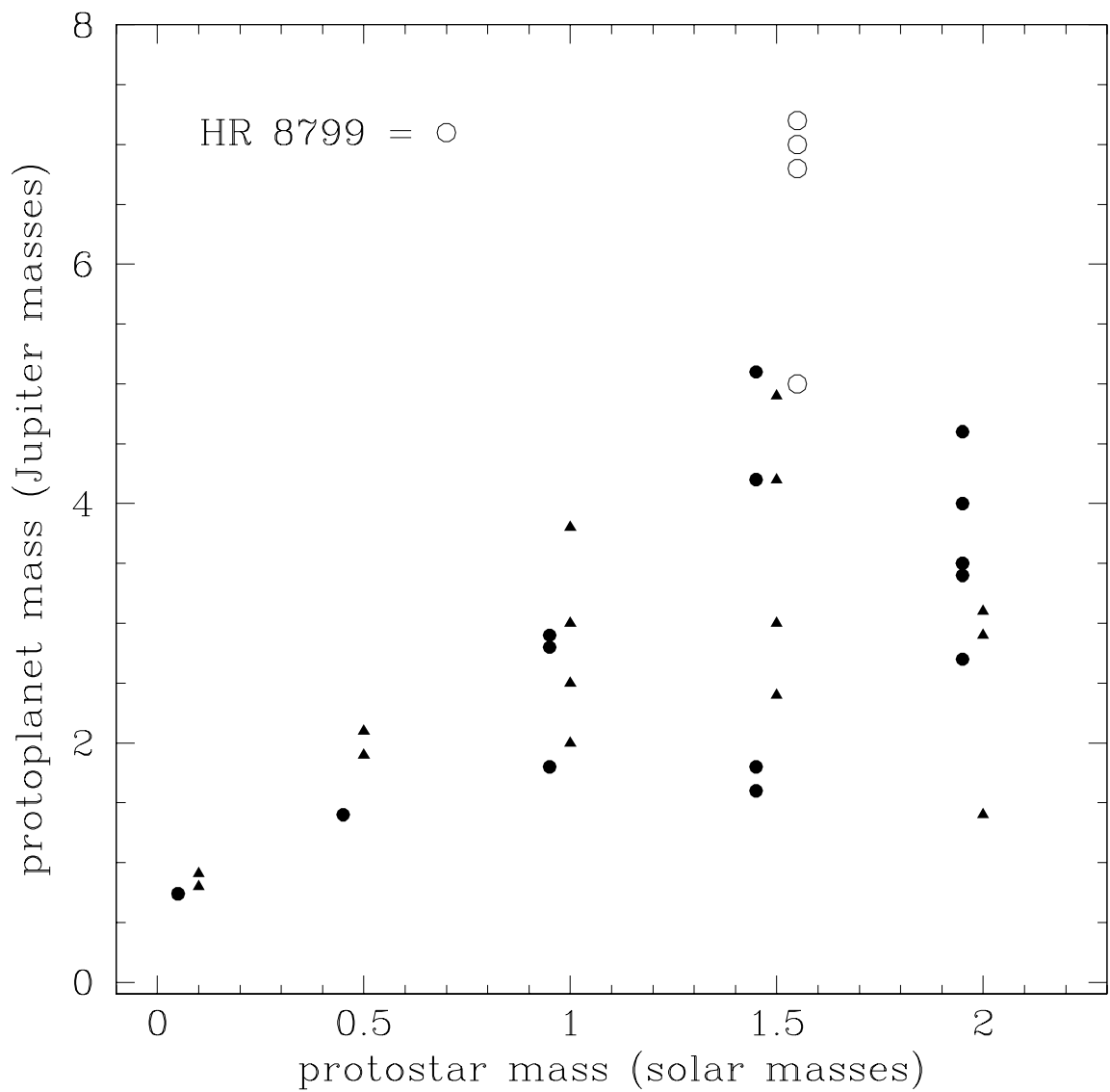}
\caption{Protoplanet masses as a function of protostellar mass, compared to
estimated planet masses for the HR 8799 system (Marois et al. 2008, 2010),
assuming an age of 30 Myr for HR 8799.
In this plot and the subsequent plots, filled circles represent estimated 
protoplanet properties at a time of $\sim 4 P_{20}$, while filled triangles 
represent the estimates at a time of $\sim 6 P_{20}$. The nominal 
protostellar masses for the filled circles have been shifted slightly
to the left for clarity, while those for HR 8799 have been shifted to 
the right, as well as up and down in mass for the three 7 $M_{Jup}$ planets.}
\end{figure}

\clearpage

\begin{figure}
\vspace{-2.0in}
\plotone{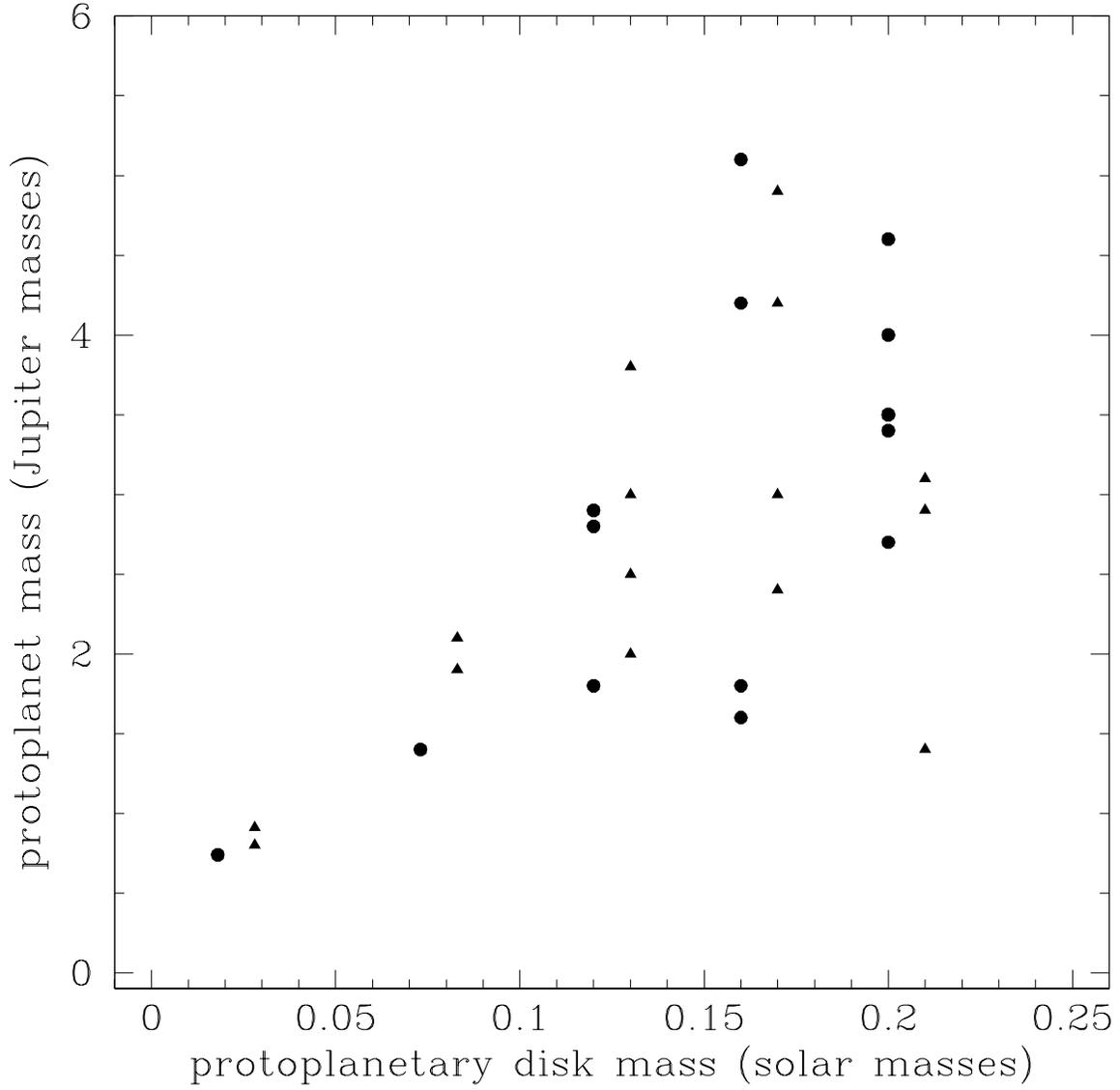}
\caption{Protoplanet masses 
as a function of protoplanetary disk mass. Filled circles have
been shifted to the left for clarity.}
\end{figure}

\clearpage

\begin{figure}
\vspace{-2.0in}
\plotone{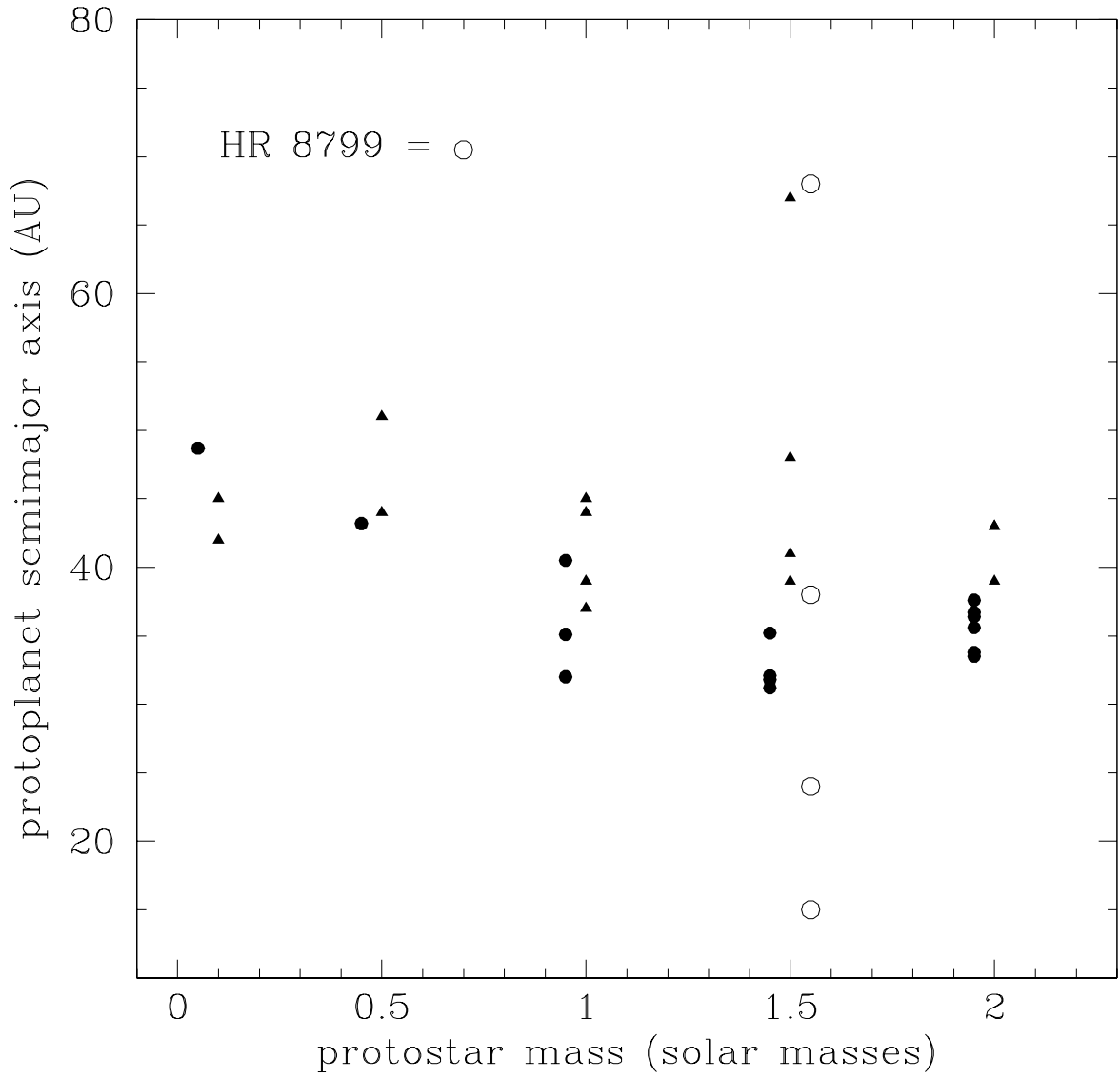}
\caption{Protoplanet orbital semimajor axes 
as a function of protostellar mass.}
\end{figure}

\clearpage

\begin{figure}
\vspace{-2.0in}
\plotone{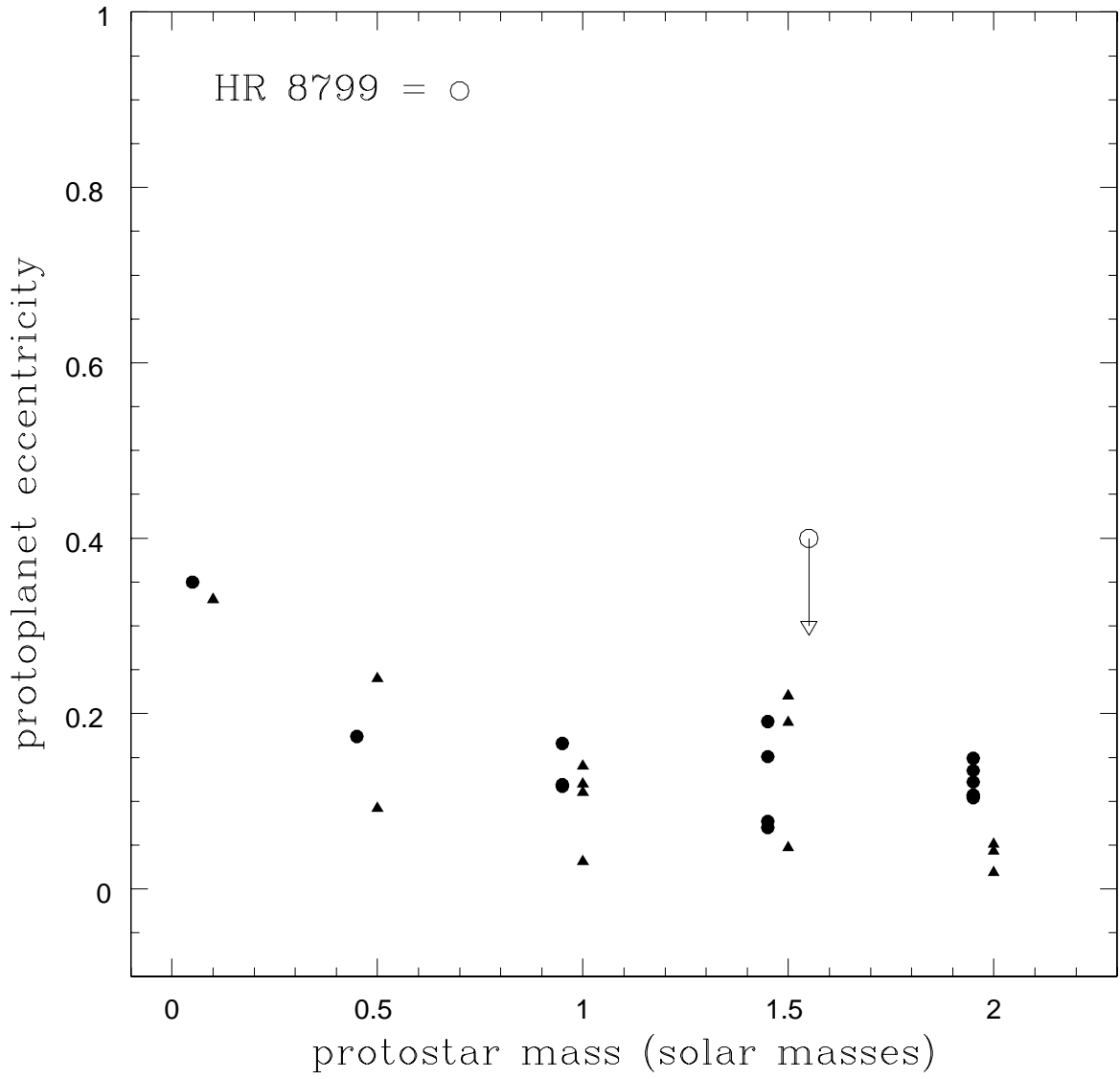}
\caption{Protoplanet orbital eccentricities 
as a function of protostellar mass.}
\end{figure}

\end{document}